\documentclass[twocolumn,showpacs,preprintnumbers,amsmath,aps]{revtex4}
\usepackage{graphicx}
\usepackage{dcolumn}
\usepackage{bm}
\usepackage{color}
\begin{document}
\def\eq#1{(\ref{#1})}
\def\fig#1{Fig.\hspace{1mm}\ref{#1}}
\def\tab#1{\hspace{1mm}\ref{#1}}
\title{Superconducting properties of tin-copper amorphous thin films}
\author{E. A. Drzazga$^{\left(1\right)}$}
\email{edrzazga@wip.pcz.pl}
\author{R. Szcz{\c{e}}{\'s}niak$^{\left(1, 2\right)}$}
\email{szczesni@wip.pcz.pl}
\author{I. A. Domagalska$^{\left(1\right)}$} 
\author{K. Piwowarska$^{\left(1\right)}$}
\affiliation{$^1$ Institute of Physics, Cz{\c{e}}stochowa University of Technology, Ave. Armii Krajowej 19, 42-200 Cz{\c{e}}stochowa, Poland}
\affiliation{$^2$ Institute of Physics, Jan D{\l}ugosz University in Cz{\c{e}}stochowa, Ave. Armii Krajowej 13/15, 42-200 Cz{\c{e}}stochowa, Poland}
\date{\today}
\begin{abstract}
The thermodynamic properties of the superconducting state in the amorphous ${\rm Sn_{1-x}Cu_{x}}$ thin films have been characterized. The concentration of copper changes in the range from $0.08$ to $0.41$. The calculations have been conducted in the framework of the strong-coupling formalism, wherein the Eliashberg functions determined in the tunnel experiment (Phys. Rev. B {\bf 51}, 685 (1995)) have been used. The value of the Coulomb pseudopotential equal to $0.1$ has been adopted. It has been found that the critical temperature ($T_{C}$) decreases from $7$ K to $3.9$ K. 
The ratio of the energy gap to the critical temperature: $R_{\Delta}=2\Delta\left(0\right)/k_{B}T_{C}$, differs significantly from 
the BCS value: $R_{\Delta}\in\left<4.4,3.95\right>$. Similarly behaves the ratio of the specific heat jump to the specific heat of the normal state: 
$R_{C}=\Delta C\left(T_{C}\right)/C^{N}\left(T_{C}\right)$, and the parameter: 
$R_{H}=T_{C}C^{N}\left(T_{C}\right)/H^{2}_{C}\left(0\right)$, where $H_{C}\left(0\right)$ is the thermodynamic critical field. 
In particular, $R_{C}\in\left<2.2,1.75\right>$ and $R_{H}\in\left<0.141,0.154\right>$.
\end{abstract}
\pacs{74.20.Fg, 74.25.Bt}
\maketitle
{\bf Keywords:} Superconductivity, Amorphous systems, Thermodynamic properties. 
\vspace*{1cm}
%


The research on the superconducting state has lasted more than hundred years \cite{Onnes1911A}. One of the first superconductors discovered was tin, wherein the superconducting phase is characterized by the low value of the critical temperature: $T_{C}=3.72$ K \cite{Eisenstein1954A}, \cite{Matthias1963A}. It turns out that the disappearance of the crystal structure and 
the formation of the amorphous system can be the reason for the significant increase in the transition temperature. In particular, the amorphous alloys containing copper deserve particular attention \cite{Schmidt1999A}. 

This paper presents the results of the theoretical analysis obtained for the superconducting state of the amorphous ${\rm Sn_{1-x}Cu_{x}}$ thin  films, for the concentration of copper in the range from $0.08$ to $0.41$. 

\begin{figure}
\includegraphics[width=\columnwidth]{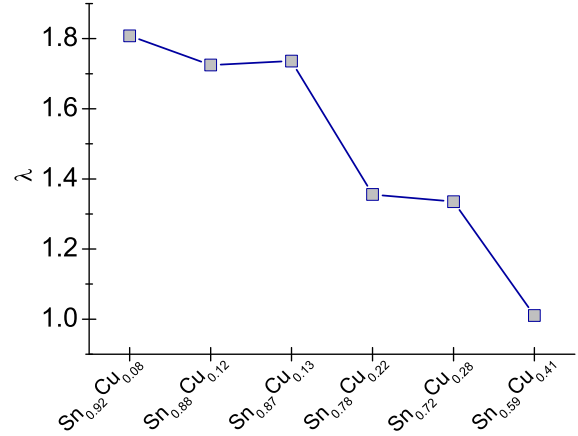}
\caption{The values of $\lambda$ as a function of the copper concentration.}
\label{f1}
\end{figure}

Due to the high values of the electron-phonon coupling constant ($\lambda\in\left<1.01,1.81\right>$), the numerical calculations have been performed in the framework of the strong-coupling theory, wherein:  
$\lambda=2\int^{\Omega_{\rm max}}_0 d\Omega \frac{\alpha^2F\left(\Omega\right)}{\Omega}$. 
The spectral functions ($\alpha^{2}F\left(\Omega\right)$) have been determined on the basis of the tunnel experiment \cite{Watson1995A}. 
The maximum phonon frequency ($\Omega_{\rm max}$) is equal to $23$ meV. A detailed dependence of the electron-phonon coupling constant on the concentration of copper is presented in \fig{f1}. 

%
%

The strong-coupling theory is based on the Eliashberg equations \cite{Eliashberg1960A}:
\begin{equation}
\label{r1}
\phi_{n}=\frac{\pi}{\beta}\sum_{m=-M}^{M}
\frac{\lambda\left(i\omega_{n}-i\omega_{m}\right)-\mu^{\star}\theta\left(\omega_{c}-|\omega_{m}|\right)}
{\sqrt{\omega_m^2Z^{2}_{m}+\phi^{2}_{m}}}\phi_{m},
\end{equation}
\begin{equation}
\label{r2}
Z_{n}=1+\frac{1}{\omega_{n}}\frac{\pi}{\beta}\sum_{m=-M}^{M}
\frac{\lambda\left(i\omega_{n}-i\omega_{m}\right)}{\sqrt{\omega_m^2Z^{2}_{m}+\phi^{2}_{m}}}\omega_{m}Z_{m}.
\end{equation}
It should be noted that the Eliashberg set can be obtained from the Fr{\"o}hlich Hamiltonian, which explicitly models the electron-phonon interaction \cite{Frohlich1952A}. The solutions of the Eliashberg equations are: the order parameter function ($\phi_{n}=\phi\left(i\omega_{n}\right)$), and the wave function renormalization factor ($Z_{n}=Z\left(i\omega_{n}\right)$). The order parameter for the superconducting state is defined as the ratio: $\Delta_{n}=\phi_{n}/Z_{n}$. The quantity $\omega_{n}$  is the fermion Matsubara frequency: $\omega_{n}=\left(\pi / \beta\right)\left(2n-1\right)$, whereas: $\beta=\left(k_{B}T\right)^{-1}$, and $k_{B}$ denotes the Boltzmann constant. The symbol $\lambda\left(z\right)$ represents the pairing kernel for the electron-phonon interaction: 
$\lambda\left(z\right)=2\int_0^{\Omega_{\rm{max}}}d\Omega\frac{\Omega}{\Omega ^2-z^{2}}\alpha^{2}F\left(\Omega\right)$.

The classical Eliashberg formalism is the one-parameter theory, in which the depairing electron correlations are modeled with the help of the Coulomb pseudopotential ($\mu^{\star}$) \cite{Morel1962A}. In the study we have assumed:  $\mu^{\star}=0.1$. The quantity $\theta$ represents the Heaviside function, $\omega_{c}$ is the cut-off frequency: $\omega_{c}=5\Omega_{\rm{max}}$.

The finite number of equations has been taken into account performing the numerical analysis ($M=1100$). The functions $\phi_{n}$ and $Z_{n}$ are stable for $T\in\left<T_{0}=1\hspace{1mm}{\rm K}, T_{C}\right>$. The Eliashberg equations have been solved using the numerical procedures tested and discussed in the papers: 
\cite{Ania2013A}, \cite{Ania2014A}, \cite{Ania2014B}, \cite{Ania2014C}, \cite{Ania2014D}, \cite{Ania2015A}, \cite{Ania2015B}.

In order to calculate the exact value of the order parameter, the solutions of the Eliashberg equations should be analytically
continued on the real axis: $\phi_{n}\rightarrow\phi\left(\omega\right)$ and $Z_{n}\rightarrow Z\left(\omega\right)$. 
Most effectively it can be done with the help of the Eliashberg equations in the mixed representation \cite{Marsiglio1988A}: 
\begin{widetext}
\begin{eqnarray}
\label{r3}
\phi\left(\omega+i\delta\right)&=&
                                  \frac{\pi}{\beta}\sum_{m=-M}^{M}
                                  \left[\lambda\left(\omega-i\omega_{m}\right)-\mu^{\star}\theta\left(\omega_{c}-|\omega_{m}|\right)\right]
                                  \frac{\phi_{m}}
                                  {\sqrt{\omega_m^2Z^{2}_{m}+\phi^{2}_{m}}}\\ \nonumber
                              &+& i\pi\int_{0}^{+\infty}d\omega^{'}\alpha^{2}F\left(\omega^{'}\right)
                                  \left[\left[N\left(\omega^{'}\right)+f\left(\omega^{'}-\omega\right)\right]
                                  \frac{\phi\left(\omega-\omega^{'}+i\delta\right)}
                                  {\sqrt{\left(\omega-\omega^{'}\right)^{2}Z^{2}\left(\omega-\omega^{'}+i\delta\right)
                                  -\phi^{2}\left(\omega-\omega^{'}+i\delta\right)}}\right]\\ \nonumber
                              &+& i\pi\int_{0}^{+\infty}d\omega^{'}\alpha^{2}F\left(\omega^{'}\right)
                                  \left[\left[N\left(\omega^{'}\right)+f\left(\omega^{'}+\omega\right)\right]
                                  \frac{\phi\left(\omega+\omega^{'}+i\delta\right)}
                                  {\sqrt{\left(\omega+\omega^{'}\right)^{2}Z^{2}\left(\omega+\omega^{'}+i\delta\right)
                                  -\phi^{2}\left(\omega+\omega^{'}+i\delta\right)}}\right],
\end{eqnarray}
and
\begin{eqnarray}
\label{r4}
Z\left(\omega+i\delta\right)&=&
                                  1+\frac{i}{\omega}\frac{\pi}{\beta}\sum_{m=-M}^{M}
                                  \lambda\left(\omega-i\omega_{m}\right)
                                  \frac{\omega_{m}Z_{m}}
                                  {\sqrt{\omega_m^2Z^{2}_{m}+\phi^{2}_{m}}}\\ \nonumber
                              &+&\frac{i\pi}{\omega}\int_{0}^{+\infty}d\omega^{'}\alpha^{2}F\left(\omega^{'}\right)
                                  \left[\left[N\left(\omega^{'}\right)+f\left(\omega^{'}-\omega\right)\right]
                                  \frac{\left(\omega-\omega^{'}\right)Z\left(\omega-\omega^{'}+i\delta\right)}
                                  {\sqrt{\left(\omega-\omega^{'}\right)^{2}Z^{2}\left(\omega-\omega^{'}+i\delta\right)
                                  -\phi^{2}\left(\omega-\omega^{'}+i\delta\right)}}\right]\\ \nonumber
                              &+&\frac{i\pi}{\omega}\int_{0}^{+\infty}d\omega^{'}\alpha^{2}F\left(\omega^{'}\right)
                                  \left[\left[N\left(\omega^{'}\right)+f\left(\omega^{'}+\omega\right)\right]
                                  \frac{\left(\omega+\omega^{'}\right)Z\left(\omega+\omega^{'}+i\delta\right)}
                                  {\sqrt{\left(\omega+\omega^{'}\right)^{2}Z^{2}\left(\omega+\omega^{'}+i\delta\right)
                                  -\phi^{2}\left(\omega+\omega^{'}+i\delta\right)}}\right]. 
\end{eqnarray}
\end{widetext}
The symbols $N\left(\omega\right)$ and $f\left(\omega\right)$ represent the Bose-Einstein and the Fermi-Dirac functions.


\vspace*{0.25cm}

%
\begin{figure}
\includegraphics[width=\columnwidth]{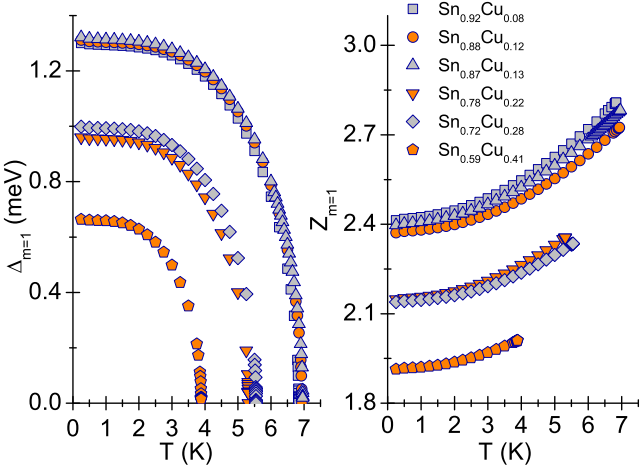}
\caption{The maximum value of the order parameter and the wave function renormalization factor in the dependence of the temperature for the selected concentration of copper.}
\label{f2}
\end{figure}
%

\fig{f2} shows the plot of the temperature course of the order parameter and the wave function renormalization factor. It can be noted that the increase in the concentration of copper very significantly weakens the superconducting state, while contributing to the less pronounced fall in the value of the electron effective mass: $m^{\star}_{e}\simeq Z_{m=1}m_{e}$, where the symbol $m_{e}$ represents the electron band mass.

%
\begin{figure}
\includegraphics[width=\columnwidth]{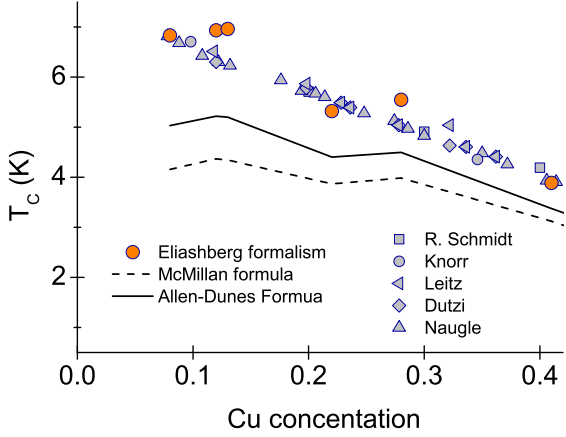}
\caption{
The influence of the copper concentration on the values of the critical temperature. The grey points represent the experimental results obtained in the papers: \cite{Schmidt1999A}, \cite{Knorr1971A}, \cite{Leitz1980A}, \cite{Dutzi1984A}, and \cite{Naugle1995A}. The black lines have been obtained with the use of the McMillan and Allen-Dynes formulas \cite{McMillan1968A}, \cite{Allen1975A}.}
\label{f3}
\end{figure}
%

Then, the overt dependence of the critical temperature on the copper concentration has been determined. The results are presented in \fig{f3}. 
The critical temperature changes over the fairly wide range from  $3.9$ K to $7$ K. Nevertheless, the full Eliashberg formalism allows us to reproduce the experimental results with a very good approximation \cite{Schmidt1999A}, \cite{Knorr1971A}, \cite{Leitz1980A}, \cite{Dutzi1984A}, \cite{Naugle1995A}. Additionally, let us point out that the values of $T_{C}$ determined with the use of the approximated McMillan or Allen-Dynes formulas are significantly understated \cite{McMillan1968A}, \cite{Allen1975A}.

The other quantities characterizing the superconducting phase (the thermodynamic critical field $H_{C}$ and the specific heat of the superconducting state $C^{S}$) have been determined on the basis of the free energy difference between the superconducting and the normal state \cite{Bardeen1964A}: 
\begin{eqnarray}
\label{r5}
\frac{\Delta F}{\rho\left(0\right)}&=&-\frac{2\pi}{\beta}\sum_{n=1}^{M}
\left(\sqrt{\omega^{2}_{n}+\Delta^{2}_{n}}- \left|\omega_{n}\right|\right)\\ \nonumber
&\times&(Z^{S}_{n}-Z^{N}_{n}\frac{\left|\omega_{n}\right|}
{\sqrt{\omega^{2}_{n}+\Delta^{2}_{n}}}).
\end{eqnarray}  
The symbol $\rho\left(0\right)$ represents the value of the electron density of states at the Fermi level. The markings $S$ and $N$ refer to the superconducting state and the normal state, respectively. 

\begin{figure}
\includegraphics[width=\columnwidth]{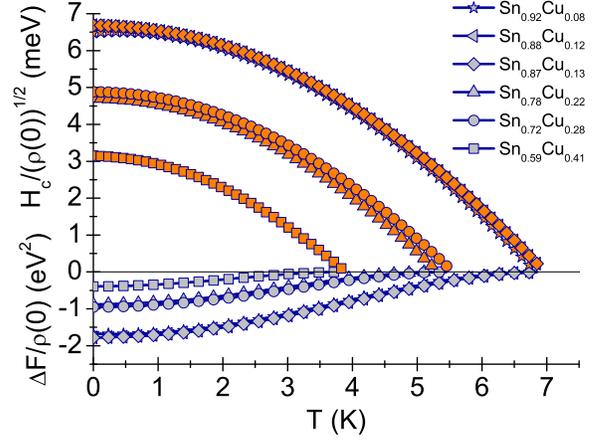}
\caption{(Lower panel) The free energy difference as a function of the temperature for the selected values of the copper concentration. (Upper panel) The dependence of the thermodynamic critical field on the temperature.}
\label{f4}
\end{figure}

The obtained curves are plotted in the lower panel in \fig{f4}. Note that an increase in the value of x causes a strong decrease of the free energy difference, which is in a direct way reflected in the thermodynamic critical field (upper panel in \fig{f4}):
\begin{equation}
\label{r6}
\frac{H_{C}}{\sqrt{\rho\left(0\right)}}=\sqrt{-8\pi\left[\Delta F/\rho\left(0\right)\right]}.
\end{equation}

%
%

The specific heat of the superconducting state should be calculated on the basis of the formula: $C^{S}=C^{N}+\Delta C$, 
where $C^{N}$ represents the specific heat of the normal state: $\frac{C^{N}\left(T\right)}{ k_{B}\rho\left(0\right)}=\frac{\gamma}{\beta}$,  
$\gamma=\frac{2}{3}\pi^{2}\left(1+\lambda\right)$ is the Sommerfeld constant. The specific heat difference between the superconducting and the normal state ($\Delta C$) is determined on the basis of:
\begin{equation}
\label{r7}
\frac{\Delta C\left(T\right)}{k_{B}\rho\left(0\right)}=-\frac{1}{\beta}\frac{d^{2}\left[\Delta F/\rho\left(0\right)\right]}{d\left(k_{B}T\right)^{2}}.
\end{equation}

\begin{figure}
\includegraphics[width=\columnwidth]{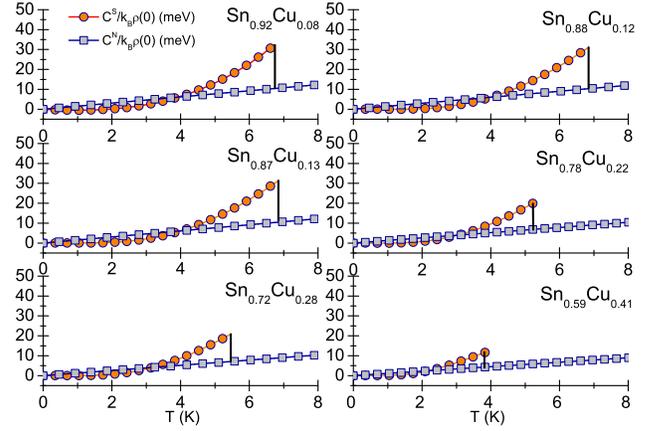}
\caption{The specific heat of the superconducting state and the specific heat of the normal state as a function of the temperature for the selected values of the copper concentration.}
\label{f5}
\end{figure}

\fig{f5} presents the dependence of the specific heat for the superconducting state and the normal state on the temperature. The characteristic jump that exists at the critical temperature is marked with the vertical line. 

%
%

%
\begin{figure*}
\includegraphics[width=2\columnwidth]{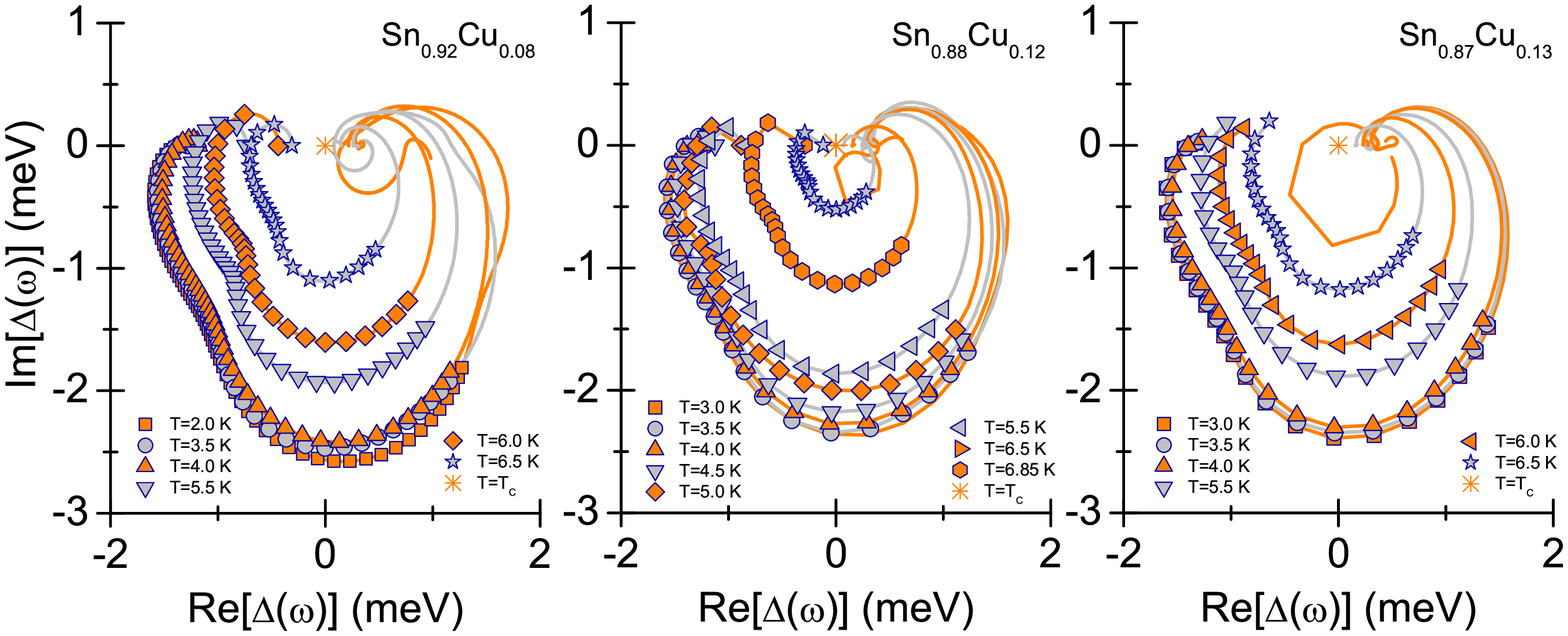}
\includegraphics[width=2\columnwidth]{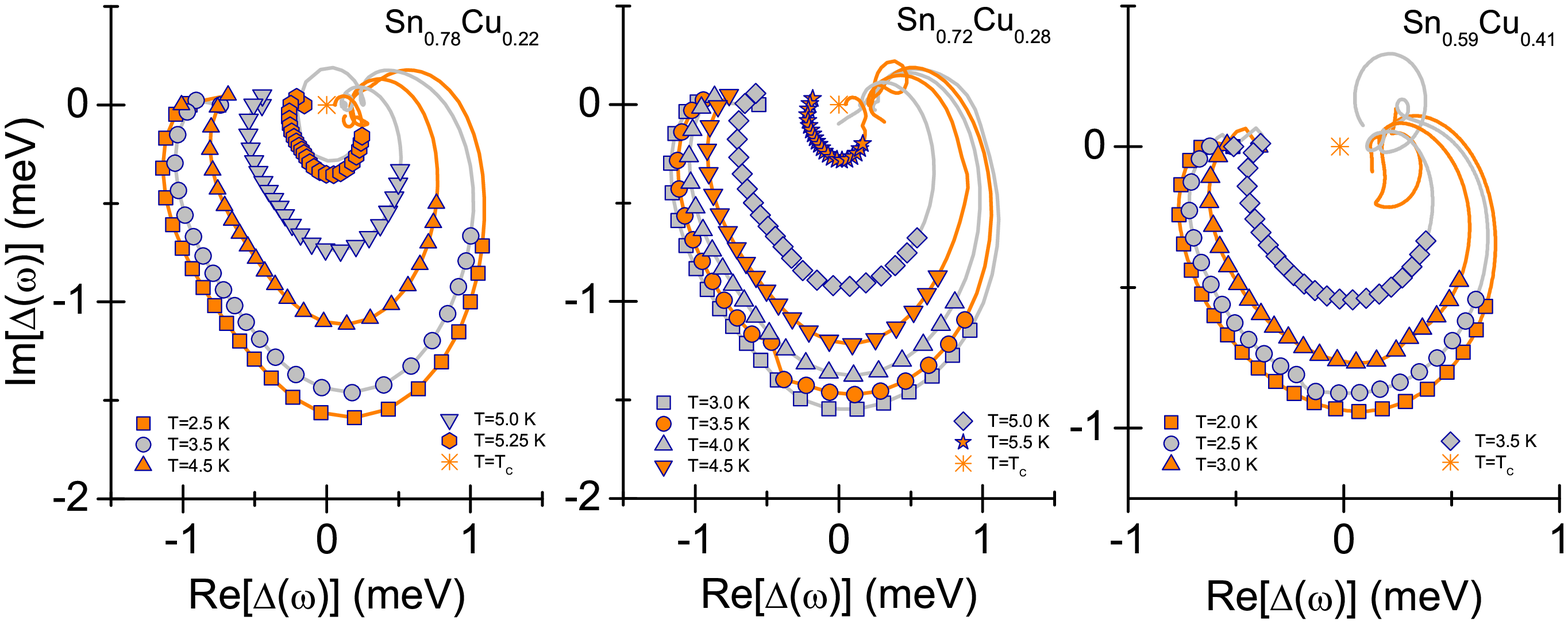}
\caption{The order parameter on the complex plane for the selected values of the temperature and the copper concentration. The lines with the symbols were obtained for $\omega\in\left<0,\Omega_{{\rm max}}\right>$. The lines without the symbols represent the frequencies from $\Omega_{{\rm max}}$ to $\omega_{c}$.}
\label{f6}
\end{figure*}
%

The analyzed thermodynamic properties allow to determine the dimensionless parameters:

\begin{equation}
\label{r8}
R_{H}=\frac{T_{C}C^{N}\left(T_{C}\right)}{H^{2}_{C}\left(0\right)}, \qquad {\rm and} \qquad 
R_{C}=\frac{\Delta C\left(T_{C}\right)}{C^{N}\left(T_{C}\right)}.
\end{equation}

The numerical calculations prove that for ${\rm x}\in\left<0.08,0.41\right>$, the parameter $R_{H}$ increases from $0.141$ to $0.154$, and the parameter $R_{C}$ decreases from $2.2$ to $1.75$. Note that the BCS theory predicts: 
$R_{H}=0.168$ and $R_{C}=1.43$ \cite{Bardeen1957A}, \cite{Bardeen1957B}. Thus, even for the high concentration of copper, the thermodynamic critical field and the specific heat of the superconducting state is not correctly described by the BCS theory. 

The physical values of the order parameter have been obtained by solving the Eliashberg equations in the mixed representation. 

\fig{f6} shows the plot of the values of the function  $\Delta\left(\omega\right)$. It can be seen that the order parameter forms the characteristic deformed spirals of the size that is decreasing together with the temperature's increase. Let us notice that the imaginary part of $\Delta\left(\omega\right)$ describes the damping effects \cite{Varelogiannis1997A}. On the other hand, the real part of the function $\Delta\left(\omega\right)$ is used to determine the exact values of the energy gap at the Fermi level: 
\begin{equation}
\label{r9}
\Delta\left(T\right)={\rm Re}\left[\Delta\left(\omega=\Delta\left(T\right)\right)\right].
\end{equation}

From the physical point of view, the most interesting is its value for the lowest temperature considered in the paper: 
$\Delta\left(0\right)=\Delta\left(T_{0}\right)$. From here, $2\Delta\left(0\right)\in\left<2.63,1.32\right>$ meV, when 
${\rm x}\in\left<0.08,0.41\right>$. 

Based on the results achieved, the values of the dimensionless parameter have been calculated:
\begin{equation}
\label{r10}
R_{\Delta}=\frac{2\Delta\left(0\right)}{k_{B}T_{C}}.
\end{equation}
As a result, it has been obtained: $R_{\Delta}\in\left<4.4,3.95\right>$. Note that in the BCS theory the ratio $R_{\Delta}$ is the universal constant of the value equal to $3.53$ \cite{Bardeen1957A}, \cite{Bardeen1957B}. 


\vspace*{0.25cm}

In conclusion, within the full Eliashberg formalism we have determined the thermodynamic parameters of the superconducting state 
in the ${\rm Sn_{1-x}Cu_{x}}$ thin films, for the concentration of copper in the range from $0.08$ to $0.41$. 

The results show that the increase in the concentration of copper lowers the value of the critical temperature from $7$ K to $3.9$ K. 
The values of the remaining thermodynamic quantities, like the order parameter, the thermodynamic critical field and the specific heat of the superconducting state, differ materially from the expectations of the BCS theory.


\bibliographystyle{ieeetr}
\bibliography{template}

%
\end{document}